\documentclass[conference]{IEEEtran}
\IEEEoverridecommandlockouts
\usepackage{amsmath,amssymb}
\usepackage{graphicx}
\usepackage{booktabs}
\usepackage{xcolor}
\usepackage{cite}
\usepackage{url}
\usepackage[hidelinks]{hyperref}
\usepackage{balance}
\graphicspath{{../figs_icc/}}
\begin{document}
\title{Capacity of Multi-Layer Manhattan-Grid UAV Corridors Served by Terrestrial 5G NR Networks}
\author{\IEEEauthorblockN{Saad Masrur, M M Sadman Shafi, Md Sharif Hossen, and Ismail Guvenc}
\IEEEauthorblockA{Department of Electrical and Computer Engineering, North Carolina State University, Raleigh, NC 27695, USA\\
Email: \{smasrur, mshafi, shossen, iguvenc\}@ncsu.edu}
\thanks{This work was supported by NASA ULI Award 80NSSC25M7102.}}
\maketitle

\begin{abstract}
Advanced air mobility (AAM) is expected to organize unmanned aerial vehicle (UAV) traffic into corridors at multiple altitudes. In this paper, we study the downlink (DL) and uplink (UL) capacity of a multi-layer Manhattan-grid corridor mesh served by a terrestrial fifth-generation (5G) new radio (NR) macro network. The corridor spacing and the numbers of layers, UAVs, and base stations are configurable. We combine the 3GPP aerial channel and antenna models, correlated UAV mobility, and frequency-selective fading with an NR physical-layer abstraction that includes channel aging and link adaptation. Results show that UAVs above 90~m experience a DL signal-to-interference-plus-noise ratio (SINR) about 17~dB below that of ground users, and that a dense macro network with 10$^\circ$ downtilt leaves 17--29\% of corridor positions without DL service under full-buffer load, regardless of corridor spacing. On a common layout, the carrier frequency (700~MHz, 3.5~GHz, or 7.2~GHz) changes the UAV SINR by less than 0.5~dB, since the corridor links are strongly interference-limited. The UL is the bottleneck: at 85\% satisfaction, a 20 MHz carrier supports only about 3.4--4.1 UAVs/km$^2$ with 4 Mb/s of UL video per UAV.
\end{abstract}


\begin{figure}[!t]
\centering
\includegraphics[width=\columnwidth]{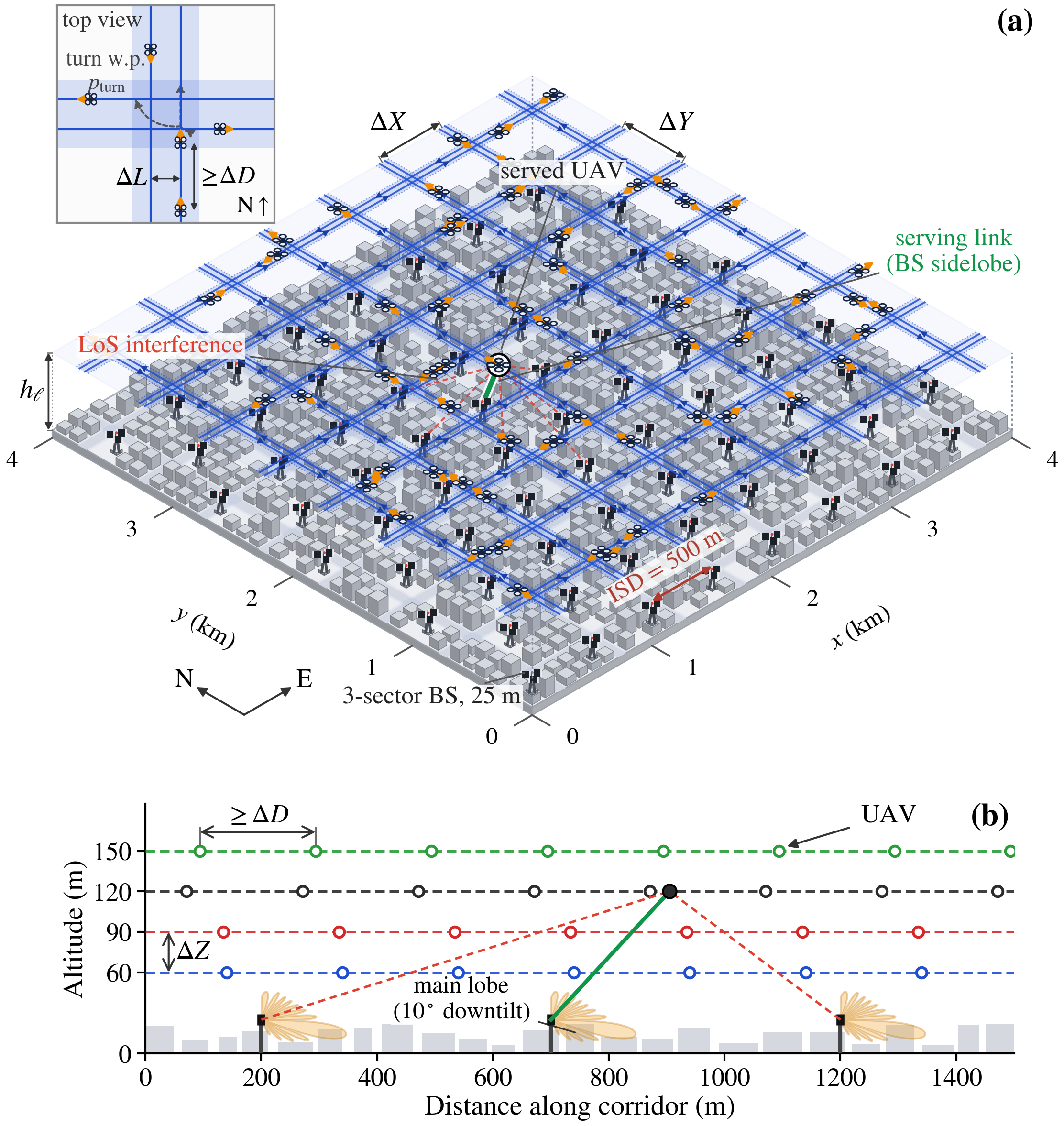}
\caption{Manhattan grid corridors scenario. (a) Perspective view of the Manhattan-grid corridors (spacing $\Delta X$, $\Delta Y$; layer altitude $h_\ell$) over three-sector BSs (25~m, ISD 500~m): the served UAV receives its serving link through a BS sidelobe and LoS interference from many BSs. The inset shows an intersection with lane separation $\Delta L$, turn probability $p_{\mathrm{turn}}$, and minimum following distance $\Delta D$. (b) Side view of the four layers at 60--150~m ($\Delta Z=30$~m), with BSs at 25~m and 10$^\circ$ downtilt.}
\label{fig:geom}
\vspace{-6mm}
\end{figure}

\section{Introduction}\label{sec:intro}
The use of unmanned aerial vehicles (UAVs) in low-altitude urban airspace is expected to grow rapidly under the advanced air mobility (AAM) concept, which organizes the airspace into corridors with internal tracks~\cite{FAA_UAM_ConOps_v2_2023}, possibly stacked in altitude layers~\cite{sunil2015metropolis, 11162478}. Every UAV in these corridors needs a reliable command-and-control (C2) link, and many also need a video link~\cite{3gpp2022ts22125}. Existing terrestrial fifth-generation (5G) networks are a natural candidate for supporting this traffic~\cite{geraci2022future}. However, their base station (BS) antennas are tilted down to serve ground users, which creates coverage holes for UAVs in corridors~\cite{maeng2023base}. Moreover, a UAV is in line of sight (LoS) with many BSs, which lowers its downlink (DL) signal-to-interference-plus-noise ratio (SINR) and spreads its uplink (UL) interference over many cells~\cite{3gpp2017tr36777}.

For a single cellular-connected UAV, these effects are well studied. A 700~MHz long-term evolution (LTE) simulation study shows that the median DL SINR of an aerial user at 40~m and 120~m is about 11~dB below that of a ground user~\cite{lin2018sky}, and field trials report median SINRs 7.7~dB and 10.2~dB below the ground-level value at 50~m and 150~m, respectively~\cite{lin2019mobile}. In the UL, a single UAV at 100~m raises the interference by up to 7.7~dB in cells within 15~km~\cite{amorim2018measured}. For UAVs that fly in corridors, our earlier work~\cite{maeng2023base} derives the corridor outage as a function of the BS uptilt; the BS tilts are optimized for corridors at 120~m and 150~m in~\cite{karimibidhendi2024optimizing}, and Bayesian optimization of the cell configuration is proposed in~\cite{benzaghta2023designing}. However, these studies consider one or a few parallel corridors at one or two altitudes and report the SINR or a Shannon rate, whereas airspace studies define layered corridor grids~\cite{sunil2015metropolis} without modeling the radio link. To the best of our knowledge, the DL and UL capacity of a multi-layer UAV corridor mesh has not been studied using a new radio (NR) physical-layer (PHY) abstraction~\cite{lagen2020new} that maps channel conditions to modulation and coding scheme (MCS)-level throughput under mobility and traffic demand.




In this paper, we study the capacity of multi-layer Manhattan-grid UAV corridors served by a 5G NR macro network as illustrated in Fig.~\ref{fig:geom}. We develop a configurable system-level model with variable corridor spacing, numbers of layers, UAVs, and BSs, and UAV mobility with correlated shadowing and LoS states. We map the SINR to throughput with an NR PHY abstraction that includes the low-rate MCS entries, channel state information (CSI) aging, and outer-loop link adaptation (OLLA). We quantify the rates, the outage, and the UL interference to ground users as functions of the corridor, traffic, and network parameters and the UAV speed. Moreover, we compare 20~MHz carriers at 700~MHz, 3.5~GHz, and 7.2~GHz and explain, through a decomposition into signal-to-noise ratio (SNR) and signal-to-interference ratio (SIR), why the carrier frequency has little effect on the UAV SINR on a common layout. Finally, with a load-coupled model, we find the UAV density that each network supports for C2 and video traffic.

\section{System Model}\label{sec:model}
Fig.~\ref{fig:geom} shows the multi-layer Manhattan-grid scenario, and Table~\ref{tab:param} shows its parameters.

\subsection{Multi-Layer Manhattan-Grid Corridors and UAV Mobility}
We consider a square area of side $W$ with wrap-around geometry, so no edge effects occur. East--west corridors lie on the lines $y=k\Delta Y$ and north--south corridors on $x=k\Delta X$, where $k$ is an integer, and each corridor has two lanes with opposite directions separated by $\Delta L$. The mesh is repeated in $L$ layers (Fig.~\ref{fig:geom}(b)) at the altitudes $h_\ell=h_0+(\ell-1)\Delta Z$, $\ell=1,\dots,L$. Each lane is divided into slots of length $\Delta D$, which represents the minimum following distance, so that the number of slots is
\begin{equation}
N_{\mathrm{slot}} = 2L\left(\frac{W}{\Delta X}+\frac{W}{\Delta Y}\right)\frac{W}{\Delta D}.
\label{eq:nslot}
\end{equation}
The $N_{\mathrm{U}}$ UAVs occupy randomly selected slots, and we define the occupancy $\zeta=N_{\mathrm{U}}/N_{\mathrm{slot}}$. The UAVs move with speed $v$ according to a multi-layer extension of the Manhattan-grid mobility model of~\cite{bai2003important}: At each intersection, a UAV continues straight with probability \(1-p_{\rm turn}\) and turns left or right with equal probability \(p_{\rm turn}/2\) while remaining in its altitude layer.


\begin{table}[!t]
\caption{Default simulation parameters (ranges in parentheses; PRB: physical resource block).}
\label{tab:param}
\centering
\footnotesize
\renewcommand{\arraystretch}{1.05}
\resizebox{\columnwidth}{!}{\begin{tabular}{@{}ll@{}}
\toprule
Parameter & Value\\
\midrule
Area; corridor spacing $\Delta X=\Delta Y$ & 4~km $\times$ 4~km; 500~m (250--2000~m)\\
Lanes, $\Delta L$; following distance $\Delta D$ & 2, 20~m; 100~m\\
Layers $L$, $h_0$, $\Delta Z$ & 4, 60~m, 30~m (1--6 layers)\\
UAVs $N_{\mathrm{U}}$; speed $v$; $p_{\mathrm{turn}}$ & 200 (50--1600); 15~m/s (5--45~m/s); 0.5\\
Sites; ISD; BS height, power & 80; 500~m (300--2000~m); 25~m, 46~dBm\\
BS array & $M=8$, $d_{\mathrm{V}}=0.8$, 10$^\circ$ downtilt\\
Carriers (20~MHz) & 0.7~GHz: 106 PRBs; 3.5, 7.2~GHz: 51 PRBs\\
Noise figure; UL power control & 5/9~dB (BS/UE); $P_0=-85$~dBm/360~kHz, $\alpha=0.8$\\
Ground UEs & 3 per cell, $\rho_{\mathrm{g}}=0.5$ (0.3 in Section~\ref{sec:demand})\\
\bottomrule
\end{tabular}}
\vspace{-6mm}
\end{table}

\subsection{Network, Antenna, and Channel Model}
The network consists of three-sector macro sites on a hexagonal grid with inter-site distance (ISD) $D_{\mathrm{ISD}}$, BS height $h_{\mathrm{BS}}$, and power $P_{\mathrm{BS}}$ per 20~MHz carrier. Each sector uses the antenna element of TR~38.901~\cite{3gpp38901} with gain $A_{\mathrm{E}}(\theta,\phi)$ in a vertical array of $M$ elements with spacing $d_{\mathrm{V}}$ wavelengths and electrical downtilt $\theta_{\mathrm{t}}$, which gives the gain
\begin{equation}
G(\theta,\phi) = A_{\mathrm{E}}(\theta,\phi) + 10\log_{10}\big(|F(\theta)|^2/M\big),
\label{eq:gain}
\end{equation}
where $F(\theta)=\sum_{m=0}^{M-1} e^{j2\pi m d_{\mathrm{V}}(\cos\theta-\cos\theta_0)}$ is the array factor, $\theta$ is the zenith angle, $\phi$ is the azimuth, and $\theta_0=90^\circ+\theta_{\mathrm{t}}$. The sidelobes of the downtilted array, which point above the horizon, serve most corridor UAVs and carry most of their interference. Each cell also serves $N_{\mathrm{g}}$ ground user equipments (UEs), whose traffic occupies a fraction $\rho_{\mathrm{g}}$ of the resources.

For the UAVs, we use the urban macro aerial-vehicle (UMa-AV) model of TR~36.777~\cite{3gpp2017tr36777}, whose LoS probability equals one above 100~m. The LoS path loss in dB is
\begin{equation}
\mathrm{PL}_{\mathrm{LoS}} = 28 + 22\log_{10} d_{\mathrm{3D}} + 20\log_{10} f_{\mathrm{c}},
\label{eq:pllos}
\end{equation}
where $d_{\mathrm{3D}}$ is the distance in meters and $f_{\mathrm{c}}$ is the carrier frequency in GHz. Ground UEs follow the TR~38.901 UMa model. The shadowing and the LoS state of each UAV--site pair evolve along the trajectory as first-order autoregressive Gaussian processes with a 50~m decorrelation distance. The LoS state is obtained by thresholding a latent Gaussian variable. The coupling gain $\mathcal{G}_{u,c}$ between cell $c$ and UE $u$ in dB is the antenna gain minus the path loss and the shadowing. Small-scale fading is Rician with a 12~dB K-factor in LoS and Rayleigh in NLoS. For the serving link and the eight strongest interferers, the fading is drawn per subband on $K_{\mathrm{sb}} = 18$ subbands of the carrier; the other interferers use the large-scale gain only.



\section{5G NR PHY Abstraction for AAM Corridors}\label{sec:phy}
Corridor UAVs operate mostly below 0~dB SINR, where the Shannon formula is a poor proxy for the NR throughput. In this section, we map the channel to the UAV throughput.

\subsection{DL and UL SINR}
Each UAV is served by the cell with the highest filtered coupling gain, with a 3~dB handover hysteresis and a 0.5~s time-to-trigger. The DL SINR of UAV $u$ on subband $k$ is
\begin{equation}
\gamma^{\mathrm{DL}}_u[k]=\frac{P_{\mathrm{BS}}\,g_{u,c_u}|H_{u,c_u}[k]|^2}{\sum_{c\neq c_u}\eta_c P_{\mathrm{BS}}\,g_{u,c}|H_{u,c}[k]|^2+N_{\mathrm{UE}}},
\label{eq:dlsinr}
\end{equation}
where $c_u$ is the serving cell, $g_{u,c}=10^{\mathcal{G}_{u,c}/10}$, $H_{u,c}[k]$ is the small-scale fading, $\eta_c\in[0,1]$ is the activity of cell $c$ ($\eta_c=1$ for full-buffer traffic), and $N_{\mathrm{UE}}$ is the noise power. In the UL, the power spectral density (PSD) of UE $u$ in dBm/Hz is
\begin{equation}
p_u=\min\left\{P_{\max}-10\log_{10}(\omega_u B),\; P_0-\alpha\,\mathcal{G}_{u,c_u}\right\},
\label{eq:pc}
\end{equation}
where $P_{\max}=23$~dBm, $\omega_u$ is the fraction of the bandwidth $B$ used by UE $u$, $P_0$ is the target PSD, and $\alpha$ is the path-loss compensation factor for UL power control. The UL interference PSD at cell $c$ is
\begin{equation}
I^{\mathrm{UL}}_c=\sum_{u\notin\mathcal{U}_c}\omega_u 10^{p_u/10} g_{u,c},
\label{eq:ulint}
\end{equation}
where $\mathcal{U}_c$ is the set of UAVs and ground UEs served by cell $c$. Evaluating~\eqref{eq:ulint} with and without UAV transmissions isolates the UL SINR loss of ground UEs due to UAV interference.


\subsection{Effective SINR and MCS Set}
The subband SINRs are mapped to an effective SINR with mutual-information effective SINR mapping (MIESM)~\cite{brueninghaus2005link},
\begin{equation}
\gamma_{\mathrm{eff},Q}=I_Q^{-1}\Big(\frac{1}{K_{\mathrm{sb}}}\sum_{k=1}^{K_{\mathrm{sb}}}I_Q(\gamma[k])\Big),
\label{eq:miesm}
\end{equation}
where $I_Q(\gamma)$ is the bit-interleaved coded modulation (BICM) mutual information of Gray-mapped quadrature amplitude modulation (QAM) with $Q$ bits per symbol.

We use the 256QAM MCS table of TS~38.214~\cite{3gpp38214}, whose lowest entry is quadrature phase-shift keying (QPSK) with code rate $R=120/1024$. Since corridor UAVs often operate below $-6$~dB, we add the six lower QPSK entries ($R=30/1024$ to $99/1024$) of the low-spectral-efficiency table, which the next-generation Node B (gNB) can select on a per-transmission basis through the MCS cell radio network temporary identifier (MCS-C-RNTI)~\cite{3gpp38214}. For MCS $m$ with modulation order $Q_m$ and code rate $R_m$, the SINR that gives 10\% block error rate (BLER) in additive white Gaussian noise (AWGN) is
\begin{equation}
\gamma^{(10\%)}_m = I_{Q_m}^{-1}(Q_m \bar{R}_m)+\Delta_{\mathrm{LDPC}}-10\log_{10}(\bar{R}_m/R_m),
\label{eq:thr}
\end{equation}
where $\bar{R}_m=\max(R_m,R_0)$, $R_0=1/5$ is the lowest code rate of low-density parity-check (LDPC) base graph~2~\cite{3gpp38212}, and $\Delta_{\mathrm{LDPC}}=1.5$~dB is the gap of a finite-length LDPC decoder to the BICM capacity. Lower rates are obtained by repetition, which gives only the energy-combining gain $10\log_{10}(R_0/R_m)$. The resulting thresholds range from $-11.8$~dB for QPSK with $R=30/1024$ (0.059~b/s/Hz) to 25.5~dB for 256QAM with $R=948/1024$ (7.41~b/s/Hz), whereas the lowest standard entry ($R=120/1024$) requires $-5.8$~dB; thus, the low-rate entries extend the coverage by about 6~dB.  Since ~\eqref{eq:thr} assumes ideal channel estimation, the lowest threshold is optimistic; increasing it by 1 dB raises the baseline DL outage from 17\% to 26\%. At the median UAV SINR of $-9$~dB, the Shannon bound gives 0.17~b/s/Hz, whereas the best MCS gives 0.088~b/s/Hz after the BLER, so the Shannon bound overestimates the UAV rate twofold.
\subsection{Link Adaptation With CSI Aging}\label{sec:la}
The MCS is selected from CSI that is $\tau_{\mathrm{CSI}}$ old, and the scattered channel components at the report and at the transmission have the correlation $J_0(2\pi f_{\mathrm{D}}\tau_{\mathrm{CSI}})$, where $f_{\mathrm{D}}=v/\lambda$ is the Doppler frequency and $\lambda$ is the carrier wavelength. At 15~m/s, the coherence time is 12~ms at 700~MHz but 1.2~ms at 7.2~GHz, so a 5~ms CSI delay leaves a correlation of 0.72 at 700~MHz and almost none at the higher carriers. The scheduler selects the highest MCS with $\hat{\gamma}_{\mathrm{eff}}-\delta_u\ge\gamma^{(10\%)}_m$, where $\hat{\gamma}_{\mathrm{eff}}$ uses the aged CSI and the OLLA offset is updated as $\delta_u\leftarrow\delta_u+\mu(\mathrm{BLER}_u-0.1)$, $\mu=2$~dB. Thus, a high BLER increases \(\delta_u\), making the next MCS more conservative. No qualifying MCS means outage.
\vspace{-0.9mm}
\subsection{Throughput and Load-Coupled Demand Model}
The rate of UAV $u$ is $r_u=\omega_u(1-\mathrm{BLER}_u)Q_{m_u}R_{m_u}B_{\mathrm{occ}}(1-\mathrm{OH})$, where $B_{\mathrm{occ}}$ is the occupied bandwidth and the overhead OH is 14\% in the DL and 8\% in the UL. With full-buffer traffic, the UAVs of a cell share the fraction $1-\rho_{\mathrm{g}}$ equally. Since a C2 link needs only about 0.1~Mb/s, we also use the load-coupled model of~\cite{siomina2012}. If UAV $u$ demands the rate $R_u$ and achieves $r^{\mathrm{full}}_u(\boldsymbol{\eta})$ when it occupies all resources of its cell, the load of cell $c$ is
\begin{equation}
\rho_c=\rho_{\mathrm{g}}+\sum_{u\in\mathcal{U}^{\mathrm{UAV}}_c}\frac{R_u}{r^{\mathrm{full}}_u(\boldsymbol{\eta})},\quad \eta_c=\min\{\rho_c,1\},
\label{eq:load}
\end{equation}
which we solve by a damped fixed-point iteration, with $\omega_u=R_u/r^{\mathrm{full}}_u$ in~\eqref{eq:ulint}. A UAV is satisfied if $R_u/r^{\mathrm{full}}_u\le 1-\rho_{\mathrm{g}}$ and $\rho_{c_u}\le 1$, and the supportable UAV density is the highest density at which a target fraction of the UAVs is satisfied in both the DL and the UL.

\section{Simulation Results}\label{sec:res}
Each full-buffer run covers 30~s of flight with a 0.5~s step and is averaged over three independent realizations; across realizations, the median SINR varies by less than 0.3~dB. The absolute DL outage is area-sensitive because larger areas include more distant LoS interferers; at 3.5~GHz, it increases from 17.3\% in 4~km $\times$ 4~km to 26.7\% in 6~km $\times$ 6~km.

\subsection{Baseline: Corridor UAVs Versus Ground UEs}
\begin{figure*}[!t]
\centering
\includegraphics[width=\textwidth]{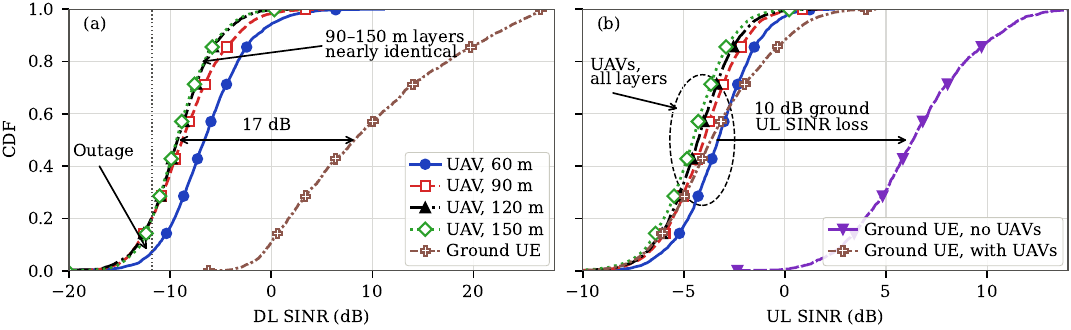}
\caption{Baseline scenario at 3.5~GHz (four layers, 200 UAVs, full buffer). (a) CDF of the DL SINR of the UAVs per layer and of the ground-UE DL geometry SINR; the dotted line marks the lowest MCS. (b) CDF of the UL SINR of the UAVs per layer and of the ground UEs with and without UAV traffic.}
\label{fig:base}
\vspace{-2mm}
\end{figure*}

In Fig.~\ref{fig:base}, we show the SINR distributions of the UAVs in the four layers and of the ground UEs. As shown in Fig.~\ref{fig:base}(a), the median DL SINR is $-6.6$~dB at 60~m and $-8.8$ to $-9.4$~dB in the upper three layers, about 17~dB below the ground-UE geometry-SINR median of 8.2~dB. The UAV distributions are narrower, since a UAV is in LoS with many cells at similar distances. As a result, 7\% of the UAVs in the lowest layer and 20--22\% in the upper layers do not qualify for the lowest MCS and are in outage, and the mean DL rate is 0.50~Mb/s per UAV. In the UL (Fig.~\ref{fig:base}(b)), the UAVs reach a median SINR of $-3.3$ to $-4.5$~dB and a mean rate of 1.17~Mb/s, while their full-buffer traffic lowers the median UL SINR of the ground UEs by 10~dB, from 6.3~dB to $-3.7$~dB.

\subsection{Effect of the UAV Altitude}
\begin{figure*}[!t]
\centering
\includegraphics[width=\textwidth]{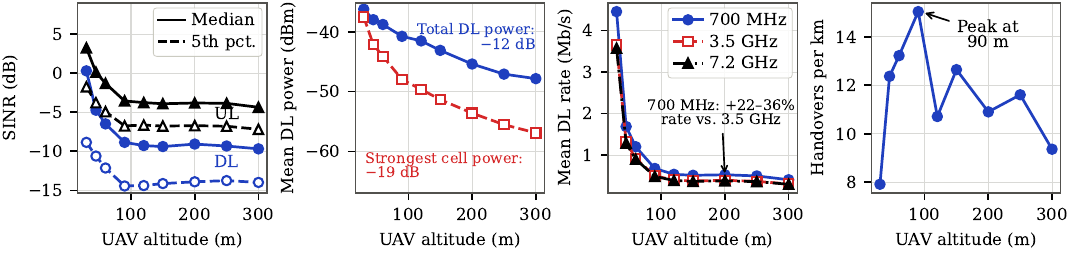}
\caption{Effect of altitude for a single layer of 150 UAVs under full-buffer traffic. (a) Median and 5th-percentile DL and UL SINR at 3.5 GHz; (b) mean total DL power and mean power from the strongest cell at the UAV; (c) mean DL rate per UAV for the three carrier frequencies; (d) handovers per km.}
\label{fig:alt}
\end{figure*}


In Fig.~\ref{fig:alt}, we show the SINR, the received power, the DL rate, and the handover rate for a single layer of UAVs as a function of its altitude. As shown in Fig.~\ref{fig:alt}(a), the median DL SINR drops from 0.3~dB at 30~m to $-8.9$~dB at 90~m and stays between $-8.9$ and $-9.7$~dB up to 300~m. Fig.~\ref{fig:alt}(b) explains this behavior: with a \(10^\circ\) downtilt, the total DL power at the UAV decreases by 12 dB from 30~m to 300~m, whereas the power from the strongest cell decreases by 19~dB. This occurs because the UAV leaves the main lobes of nearby BSs and is increasingly served through sidelobes. Above 100 m, all links are in LoS, and the sidelobe losses of the serving and interfering cells largely cancel, so the SINR no longer drops significantly. The mean DL rate in Fig.~\ref{fig:alt}(c) falls from 3.7~Mb/s at 30~m to about 0.4~Mb/s above 90~m at 3.5~GHz. The handover rate in Fig.~\ref{fig:alt}(d) peaks at 15 per km at 90 m, where the UAV crosses the sidelobes of many BSs, so a UAV at 15 m/s hands over every 4.4 s at this altitude.

\subsection{Effect of the Number of Layers and UAVs}
\begin{figure*}[!t]
\centering
\includegraphics[width=\textwidth]{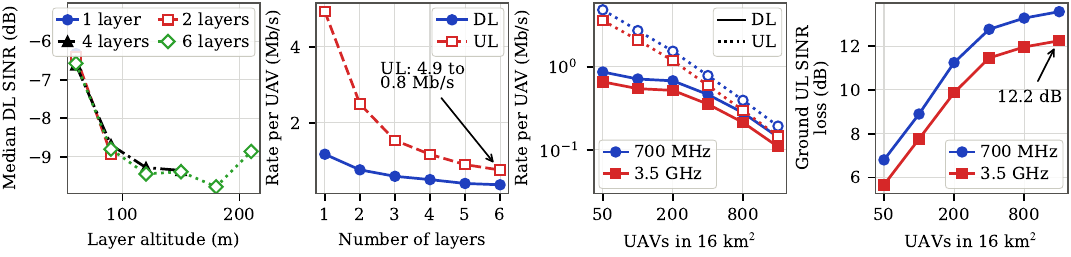}
\caption{Effect of UAV traffic under full-buffer operation. (a) Median DL SINR by layer altitude and (b) mean DL/UL rate per UAV as the number of layers increases from one to six, with 50 UAVs per layer at 3.5 GHz (\(h_0=60\) m, \(\Delta Z=30\) m); (c) mean DL/UL rate per UAV and (d) ground-UE UL SINR loss versus total UAVs in four layers at 700 MHz and 3.5 GHz.}
\label{fig:traffic}
\vspace{-6mm}
\end{figure*}

 In Fig.~\ref{fig:traffic}, we show the effect of the number of layers and UAVs. As shown in Fig.~\ref{fig:traffic}(a), the DL SINR of a layer depends only on its altitude, since the DL interference comes from the BSs, not from the UAVs. The small rise at 210~m is caused by the sidelobe pattern of the vertical BS array. The UL, in contrast, degrades with every added layer as each layer introduces 50 more UAVs. From one to six layers, the mean UL rate per UAV falls from 4.9 to 0.8~Mb/s (Fig.~\ref{fig:traffic}(b)) and the UL SINR loss of the ground UEs grows from 4.6 to 10.6~dB, since each added UAV raises the UL interference in every cell in its LoS. More UAVs have a similar effect: at 3.5~GHz, from 50 to 1600 UAVs in 16~km$^2$, the mean DL rate per UAV decreases from 0.66 to 0.11~Mb/s (Fig.~\ref{fig:traffic}(c)), the aggregate UAV throughput saturates at about 11~Mb/s/km$^2$ (DL) and 15~Mb/s/km$^2$ (UL), and the ground-UE UL SINR loss reaches 12.2~dB (Fig.~\ref{fig:traffic}(d)). The DL outage stays at 16--17\%, as full-load DL interference is independent of the number of UAVs.


\subsection{Effect of Corridor Spacing and the Network Deployment}
\begin{figure*}[!t]
\centering
\includegraphics[width=\textwidth]{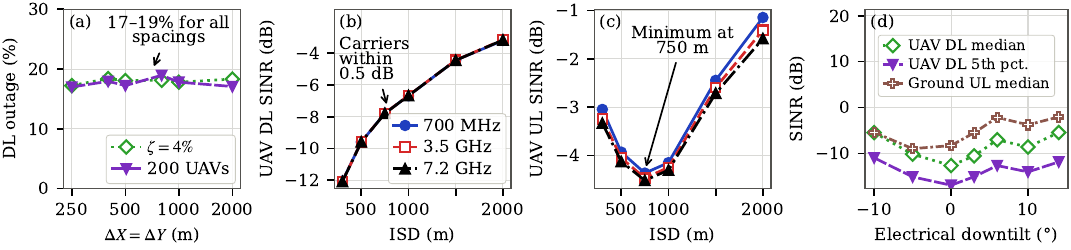}
\caption{Effect of corridor spacing and network configuration (full buffer). (a) DL outage at 3.5~GHz versus corridor spacing, at occupancy $\zeta=4\%$ and with 200 UAVs; (b) median UAV DL and (c) UL SINR versus ISD for the three carriers (300 UAVs, 6~km $\times$ 6~km); (d) UAV DL SINR and ground-UE UL SINR versus electrical tilt at 3.5~GHz, where negative values denote uptilt.}
\label{fig:net}
\vspace{-3mm}
\end{figure*}

In Fig.~\ref{fig:net}, we show the effect of corridor spacing, ISD, and antenna tilt. As the corridor spacing increases from 250~m to 2000~m, the number of available corridor slots decreases. Thus, at fixed occupancy $\zeta=4\%$, the UAV population falls from 410 to 51, reducing UL interference and increasing the mean UL rate per UAV from 0.59 to 3.2~Mb/s (not shown). For both fixed occupancy and 200 UAVs, the DL outage remains nearly unchanged at 17--19\% (Fig.~\ref{fig:net}(a)), showing that corridor spacing itself has little effect on DL coverage. Network densification has a much stronger impact: as the ISD decreases from 2000~m (36 cells) to 300~m (1440 cells), the median DL SINR drops from $-3.2$ to $-12.1$~dB (Fig.~\ref{fig:net}(b)) and the DL outage rises from 0.8\% to 53\% (not shown), because the additional LoS BSs mainly increase interference rather than serving-cell power. At every ISD, the three carriers remain within 0.5~dB in SINR. The UL SINR in Fig.~\ref{fig:net}(c) is lowest at an ISD of 750~m: small cells give a high received power under power control, while large cells have fewer interfering UEs in LoS. Antenna tilt also has a strong effect (Fig.~\ref{fig:net}(d)): the DL outage is 2.7\% with a 10$^\circ$ uptilt, 61\% with no tilt, 17\% with the default 10$^\circ$ downtilt, and 5\% with a 14$^\circ$ downtilt. The stronger downtilt suppresses interference toward the corridors, highlighting a tradeoff between UAV and ground-user performance~\cite{maeng2023base,karimibidhendi2024optimizing,benzaghta2023designing}.


\subsection{Band-Specific Deployments}\label{sec:prof}
\begin{table}[!t]
\caption{Deployment profiles and full-buffer results (200 UAVs, four layers, 36~km$^2$).}
\label{tab:prof}
\centering
\footnotesize
\renewcommand{\arraystretch}{1.05}
\resizebox{\columnwidth}{!}{\begin{tabular}{@{}lrrr@{}}
\toprule
 & 700~MHz & 3.5~GHz & 7.2~GHz\\
\midrule
ISD (m); number of cells & 1500; 72 & 500; 504 & 400; 810\\
Array size $M$, downtilt & 4, 6$^\circ$ & 8, 10$^\circ$ & 16, 10$^\circ$\\
\midrule
Median serving power (dBm) & $-36.8$ & $-50.8$ & $-59.2$\\
Median SNR / SIR (dB) & 55.4 / $-6.5$ & 41.6 / $-9.4$ & 33.2 / $-9.7$\\
DL SINR, median / 5th pct. (dB) & $-6.5$ / $-11.5$ & $-9.5$ / $-15.2$ & $-9.8$ / $-15.7$\\
DL outage (\%) & 3.8 & 26.7 & 29.4\\
UL SINR, median (dB) & $-3.9$ & $-2.8$ & $-1.8$\\
Mean DL / UL rate (Mb/s) & 0.58 / 0.77 & 0.54 / 1.86 & 0.49 / 2.36\\
Handovers per km & 7.8 & 11.8 & 16.8\\
\bottomrule
\end{tabular}}
\vspace{-3mm}
\end{table}

Since actual deployments differ by band, we define the profiles in Table~\ref{tab:prof}: a sparse 700~MHz network with a short array, the baseline 3.5~GHz network, and a denser 7.2~GHz network with a 16-element array of the same physical height. The 3.5 and 7.2~GHz networks give nearly the same DL SINR, with medians of $-9.5$ and $-9.8$~dB, whereas the sparse 700~MHz network gives $-6.5$~dB and only 3.8\% outage. The higher outage than in Fig.~\ref{fig:base}(a) is due to the larger 6~km $\times$ 6~km area, which includes more distant LoS interferers. The SNR and SIR rows explain why the higher path loss at 7.2~GHz does not lead to much lower SINR: although its median serving power is 8~dB below that at 3.5~GHz, the median SNR is still 33~dB, while the median SIR is about $-9.5$~dB. Thus, the UAV links are strongly interference-limited. The SNR and SIR in Table~\ref{tab:prof} are computed from the mean received powers without small-scale fading, while the SINR is the subband-averaged SINR with fading; all are evaluated on the same samples. Since the serving and interfering LoS links follow the same path-loss law, the frequency term in~\eqref{eq:pllos} largely cancels in their power ratio. The denser 7.2~GHz deployment shortens the serving distance, but also adds more nearby LoS interferers, so its DL SINR remains close to that at 3.5~GHz. In contrast, the sparse 700~MHz network has fewer LoS interferers and therefore a better SIR, so its DL advantage comes mainly from the deployment, including density and array configuration, rather than carrier frequency. The UL shows the opposite trend: the denser 7.2~GHz network gives the highest UL rate because smaller cells reduce UAV transmit power and inter-cell interference, but it requires 16.8 handovers per km.

\begin{figure}[!t]
\centering
\includegraphics[width=\columnwidth]{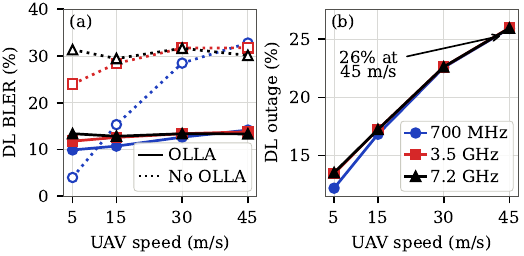}
\caption{Effect of the UAV speed (baseline, 200 UAVs, full buffer). (a) DL BLER with and without OLLA (5~ms CSI delay). (b) DL outage with OLLA.}
\label{fig:speed}
\end{figure}
\begin{figure}[!t]
\centering
\includegraphics[width=\columnwidth]{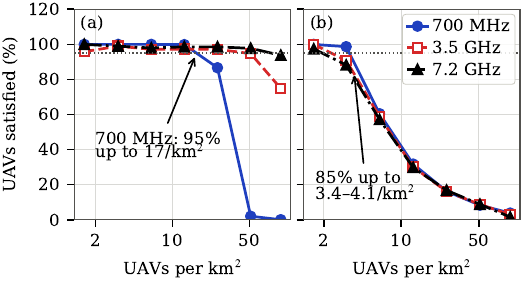}
\caption{Fraction of UAVs with both DL and UL demands met versus UAV density (load-coupled model, four layers, $\Delta X=500$~m). (a) Deployment profiles of Table~\ref{tab:prof}, C2 traffic (0.1~Mb/s DL and UL). (b) Common layout (ISD 500~m, 10$^\circ$ downtilt), C2 plus 4~Mb/s UL video. The dotted line marks the 95\% target.}
\label{fig:cap}
\vspace{-3mm}
\end{figure}

\subsection{Effect of the UAV Speed}\label{sec:speed}

In Fig.~\ref{fig:speed}, we show the DL BLER and outage versus the UAV speed. As shown in Fig.~\ref{fig:speed}(a), without OLLA, the BLER of the served UAVs is 24--32\% at 3.5 and 7.2~GHz at all speeds, since the CSI is outdated after 5~ms; at 700~MHz, it grows from 4\% at 5~m/s to 33\% at 45~m/s. With OLLA, the BLER stays at 10–14\% because OLLA backs off the MCS (Section~\ref{sec:la}). A 1~ms CSI delay (not shown) recovers most of the rate at 700~MHz, while the coherence time falls below 1~ms above about 36~m/s at 3.5~GHz and 18~m/s at 7.2~GHz. This slower aging gives the 700~MHz carrier its 22--36\% higher rate in Fig.~\ref{fig:alt}(c). The DL outage in Fig.~\ref{fig:speed}(b) increases from 12--14\% at 5~m/s to 26\% at 45~m/s for all carriers because of the handover delay: the handover lags by 0.5--1~s, during which a fast UAV can move from a serving-cell sidelobe into a null.

\subsection{Supportable UAV Density}\label{sec:demand}
We now replace full-buffer traffic with two demand profiles: C2 with 0.1~Mb/s in both the DL and UL~\cite{3gpp2017tr36777}, and C2 plus a 4~Mb/s UL remote-pilot video stream~\cite{3gpp2022ts22125}. Fig.~\ref{fig:cap}(a) shows the band-specific deployments under C2 traffic. Since the C2 demand is small, all three networks satisfy nearly all UAVs at low density. Unlike in the full-buffer case, the light C2 load reduces the cell activity $\eta_c$ in~\eqref{eq:dlsinr} and thus the DL interference, so the DL outage nearly vanishes. As the density increases, however, the sparse 700~MHz deployment becomes load-limited first because its larger cells must serve more UAVs. It supports about 17 UAVs/km$^2$ at a 95\% satisfaction target, compared with about 48 UAVs/km$^2$ at 3.5~GHz and 79 UAVs/km$^2$ at 7.2~GHz. The denser 3.5 and 7.2~GHz deployments distribute the UAV load over more cells and therefore support higher C2 densities.

Adding 4~Mb/s UL video changes the limiting mechanism. On the common layout in Fig.~\ref{fig:cap}(b), the per-UAV UL demand increases by about 40$\times$, so UL capacity and interference become the bottleneck. The three carriers therefore show similar behavior, supporting about 4.1, 3.6, and 3.4 UAVs/km$^2$ at an 85\% satisfaction target for 700~MHz, 3.5~GHz, and 7.2~GHz, respectively. These densities correspond to about 14--17~Mb/s/km$^2$ of aggregate UL demand per 20~MHz carrier. The similar results across the three bands align with the common-layout SINR results: once the deployment geometry is fixed, UAV links are strongly interference-limited, while higher video demand dominates the supported density.


\section{Conclusion}\label{sec:conc}
In this paper, we evaluated the DL and UL capacity of multi-layer Manhattan-grid UAV corridors under a 5G NR macro deployment with NR link adaptation, mobility, and load coupling. Above 90~m, UAVs are served mainly through BS sidelobes and see about a 17~dB lower DL SINR than ground UEs. Corridor spacing has little effect on DL coverage, whereas BS density and antenna tilt strongly affect outage; a 10$^\circ$ uptilt reduces the DL outage from 17\% to 2.7\%, at a cost to ground users. On a common layout, the SINR differs by less than 0.5~dB from 700~MHz to 7.2~GHz, consistent with an interference-limited regime, and the carrier matters mainly through channel aging. C2 traffic is supported up to 17--79 UAVs/km$^2$ at 95\% satisfaction, but with 4~Mb/s of UL video per UAV, a 20~MHz carrier supports only 3.4--4.1 UAVs/km$^2$ at 85\%, making the UL the bottleneck. These results assume omnidirectional UAV antennas, ideal channel estimation in the MCS thresholds, and a finite wrap-around area. Future work will consider massive MIMO and dedicated aerial beams.
\balance
\section*{Acknowledgment}
The authors used Anthropic's Claude Code and Claude Cowork to assist with simulation code development, data analysis, and drafting and editing of the manuscript text. All AI-assisted content was independently verified by the authors. The authors take full responsibility for the work.

\bibliographystyle{IEEEtran}
\bibliography{refs}
\end{document}